\documentclass[pra,amssymb,amsfonts,twocolumn,superscriptaddress,showpacs]{revtex4}
\usepackage{amsmath}
\usepackage{amsfonts}
\usepackage{amssymb}
\usepackage{bm}
\usepackage{epsfig}
\def\beq{\begin{equation}}
\def\eeq{\end{equation}}
\def\0{\otimes}
\def\1{\mbox{1\hskip-.25em l}}  
\def\6{\langle}
\def\9{\rangle}
\def\tr{{\rm tr}}
\def\pad{\partial}

\def\half{\mbox{$1\over2$}}
\def\quart{\mbox{$1\over4$}}

\def\bb{{\bf b}}

\def\bk{{\bf k}}
\def\bp{{\bf p}}
\def\br{{\bf r}}
\def\bx{{\bf x}}
\def\by{{\bf y}}
\def\bz{{\bf z}}

\def\bA{{\bf A}}
\def\bE{{\bf E}}
\def\bB{{\bf B}}

\def\bep{\mbox{\boldmath $\epsilon$}}
\def\bal{\mbox{\boldmath $\alpha$}}
\def\hbphi{\hat{\mbox{\boldmath $\phi$}}}

\def\hbk{\hat{\bk}}
\def\hbr{\hat{\br}}
\def\hbj{{\textbf{\^{\j}}}}
\def\hbx{\hat{\bx}}
\def\hby{\hat{\by}}
\def\hbz{\hat{\bz}}

\def\rmm{{\rm m}}
\def\vac{{\rm vac}}

\def\bay{\begin{array}}
\def\eay{\end{array}}
\begin{document}
\title{The effect of focusing on polarization qubits}

\author{Netanel H. Lindner}\email{lindner@techunix.technion.ac.il}
\affiliation{Department of Physics, Technion --- Israel Institute of
Technology,  Haifa 32000, Israel}

\author{Daniel R. Terno}\email{dterno@perimeterinstitute.ca}
\affiliation{Perimeter Institute for Theoretical Physics,
Waterloo N2J 2W9, Ontario, Canada }

\begin{abstract}
 When a photon with well-defined
polarization and momentum passes through a focusing device, these
properties are  no longer  well defined. Their loss is captured by
describing  polarization by a $3\times 3$ effective density
matrix.  Here we show that the effective density matrix
corresponds to the actual photodetection model and we provide a
simple formula to calculate it in terms of classical fields.
Moreover, we explore several possible experimental consequences of
the ``longitudinal" term: limits on single-photon detection
efficiency, polarization-dependent atomic transitions rates and
the implications on quantum information processing.

\end{abstract}
\maketitle

\section{Introduction}

Single-photon manipulations have recently become an integral part
of quantum optics and  they play an important role in experimental
quantum information processing \cite{exam}. Already with  current
technology the single-photon states can be produced and
manipulated quite reliably ~\cite{uren:03}, and the efficiency of
their production and detection is expected to rise in the future.
Usually these states are considered to be  eigenstates of
momentum, and often the very notion of photon is synonymous with
an elementary excitation of the electromagnetic field with a
well-defined momentum and polarization. A single-photon state
$|\bk,\sigma\9=\hat{a}^\dag_{\bk\sigma}|\vac\9$  is  an excitation
of the plane wave mode of momentum $\bk$ and polarization $\sigma$
and is distributed over the entire space.
 An
approximately localized photon
\cite{mandel} is described instead by a superposition
\beq
|\Psi\9=\int d\mu(\bk)\sum_{\sigma}f_{\sigma}(\bk)|\bk,\sigma\9,
\eeq
where $\sigma$ denotes the helicity, the normalization is given by
$\sum_\sigma\int d\mu(k)|f_\sigma(\bk)|^2$ $=1$, while  we adopt
 a non-relativistic measure $d\mu(\bk)=d^3\bk/(2\pi)^3$.
 More generally,  localized states may be described as  mixtures of such
 terms.

 A typical spread in momentum
 is often very small, and polarization is approximately constant.
As a result, polarization is usually described by a $2\times 2$
reduced density matrix which is formally equivalent to that of a
qubit, the ideal unit of quantum information
\cite{ap:b}. To ensure  validity of this approximation and to
calculate possible corrections, a general notion of polarization
density matrix is required. However, the standard definition of
reduced density matrix fails for photon polarization \cite{lpt},
and it is possible to define only an effective $3\times 3$
density matrix which corresponds to a restricted class of positive
operator-valued measures
\cite{pt:03,rmp}. Both experemental results and theoretical calculations for
classical fields and for quantum coherent states
\cite{dorn:03,enk,enk:03} establish new effects that follow from
the presence of a significant longitudinal electric field. This is
often created by using focusing devices, which are also an
inevitable part of usual optical experiments.

In this paper we apply the effective density matrix formalism to
quantum states that correspond to classical modes with a
significant longitudinal component and discuss its connection with
experiment. The paper is organized as follows. Sec.~II reviews
effective $3\times 3$ density matrices and describes some of their
applications.  Sec.~III  presents a general formula that expresses
an effective  density matrix in terms of classical modes. It is
illustrated by an elementary discussion of the lens action on
one-photon states. Sec.~IV discusses a connection between the
formal construction of effective density matrix and a simple
photodetection model. Finally, Sec.~V  introduces possible
experimental consequences of the ``longitudinal" term in effective
density matrices.

\section{Effective density matrix and its applications}

If one is interested only in polarization degrees of freedom, it
is tempting to define a reduced $2\times 2$ density matrix by
\beq \rho_{\sigma,\sigma'}=\int
d\mu(\bk)f_{\sigma}(k)f^*_{\sigma'}(k). \label{badreduced}
\eeq
However,  helicity eigenstates are  defined only with respect to a
given momentum. Intuitively, the polarization vectors for
different momenta lie in different planes and cannot be
superimposed. Under rotations each component acquires a
momentum-dependent phase and hence the density matrix
(\ref{badreduced}) has no definite transformation properties
\cite{lpt,rmp}. This makes a standard density matrix a useless
concept even when  a fixed reference frame is considered, since
any POVM (positive operator-valued measure
\cite{ap:b,bus:95}) that describes an experimental setup must have
definite transformation properties at least under ordinary
rotations. Analysis of the one-photon scattering
\cite{aie} gives another angle on  the failure of this
concept.

The $3\times 3$ matrix with the right transformation properties
may be introduced with the help of  polarization 3-vectors
\cite{rmp}.
 A polarization state
$|\bal(\bk)\9$ corresponds to the geometrical 3-vector
\beq
\bal(\bk)=\alpha_+(\bk)\bep^+_\bk+\alpha_-(\bk)\bep^-_\bk,
 \label{elliptic}
 \eeq
where $|\alpha_+|^2+|\alpha_-|^2=1$, and the vectors
$\bep^\pm_\bk$ correspond to the right and left circular
polarization and satisfy $\bk\cdot\bep^\pm_\bk=0$. In this
 notation a generic one-photon state can be written as
\beq
|\Psi\9=\int d\mu(\bk)f(\bk)|\bk,\bal(\bk)\9\label{photon},
\eeq
and the effective $3\times 3$ description takes the form
\beq
\rho_{mn}=\int d\mu(\bk)|f(\bk)|^2\bal_m(\bk)\bal_n(\bk)^*,
\label{reduce}
\eeq
where $m,n=x,y,z$ and the vector $\bal(\bk)$ is given by Eq.~
(\ref{elliptic}). Under  rotations of the coordinate system this
density matrix has a simple transformation law,
\beq
\rho \rightarrow R \rho R^T,
\label{rotation}
\eeq
where $R$ is the  rotation matrix. The effective density matrix
$\rho$ can be obtained with the standard state-reconstruction
techniques from a family of POVMs \cite{pt:03}. For example, its
diagonal elements $\rho_{ii}$ are  the expectation values of the
elements of the ``momentum-independent" polarization POVM  that
consists of three positive operators $E_i$ that sum up to the
identity, $\sum_i E_i=\1$,
\beq
\rho_{ii}=\6\Psi|E_i|\Psi\9. \label{POVM0}
\eeq
We discuss a relation between this POVM and the standard
photodetection model in  Sec.~IV.

Let us now examine a wave packet which describes a nearly plane
and nearly monochromatic wave. In this limit, $f(\bk)$ of
Eq.~(\ref{photon}) is strongly localized around some central value
$\bk_0$.  By an appropriate transformation of the form
(\ref{rotation}) the effective $3\times 3$ reduced density matrix
can be put  into a block diagonal form. Then the density matrix
will have a $2\times 2$ block with almost unit trace, which
 corresponds to the standard $2\times 2$ reduced
density matrix that would describe the state for
$f(\bk)\propto\delta(\bk-\bk_0)$. As an example, consider the
following wavepackets that are formed by the helicity eigenstates,
\beq
|\Psi_\pm\9=\int d\mu(\bk) f(\bk)|\bk,\bep^\pm_\bk\9\label{st},
\eeq
where $f(\bk)$ satisfies the above criteria. Calculating the
effective reduced density matrix we have
\beq
\rho_+=\half(1-\half\Omega^2)\left(\bay{ccc}
1 & -i & 0\\
i & 1 &0\\
0 & 0 &0
\eay\right)+\half\Omega^2\left(\bay{ccc}
0 & 0 & 0\\
0 & 0 & 0\\
0 & 0 & 1
\eay\right),
\eeq
where $\Omega\ll 1$ is roughly the ratio of a typical
 width of the wave packet  to $|\bk_0|$, and $\rho_-=\rho_+^*$.
 Its exact form depends on the detailed shape of
$f(\bk)$. Thus we arrive to the conclusion that integrating out
the photon's momentum leads to polarization states that are
neither pure nor perfectly distinguishable
\cite{lpt}. There is
 a non-zero probability to identify a $\rho_+$ state as a
$\rho_-$ state and vice versa. The probability of making such an
error given a perfect equipment and measurement scheme
\cite{fuchs} will be denoted by $P_E$ and is given by \cite{rmp}
\beq
P_E(\rho_+,\rho_-)=\half-\quart \tr |\rho_+ -
\rho_-|\approx\half\Omega^2
\label{PE}
\eeq
Although conceptually the implications of the above conclusions
are profound,
 when considering a wave packet with a very narrow
distribution in momentum $f(\bk)$, the  parameter $\Omega$ is very
small. Consider an electromagnetic beam propagating along the
$z$-axis and with the Gaussian distribution in intensity in the
$(xy)$ plane that is given by $I(r)=I_0
\exp(-r^2/\tau^2)$. Since the distribution in momentum $f(\bk)$
is essentially the Fourier transform of the electric field, we
expect that the radial spread in momentum will also be of the
Gaussian form $f(\bk)\propto f_1(k_z)
\exp(-k_r^2/2\Delta_r^2)$ where $\Delta_r\sim 1/\tau$, and $f_1$
is some function of $k_z$. Assuming also the Gaussian distribution
in wavelength $f_1(k_z)\sim\exp(-(k_z-k_0)^2/2\Delta_z^2)$ we have
\beq
f(\bk)= N \exp(-(k_z-k_0)^2/2\Delta_z^2)
\exp(-k_r^2/2\Delta_r^2).
\label{numbers}
\eeq
For this distribution in momentum, the parameter $\Omega$ is given
by \cite{lpt} $\Omega=\Delta_r/k_0 +O(\Delta_r^2/k_0^2)$. Taking
$\tau$ to be of the orders of $10^{-3}\rm m$, and a wavelength of
$5\times10^{-7}\rm m$, we get $\Omega\sim 5\times10^{-4}$, thus
rendering the effect negligible.

However, the  momentum spread becomes substantial when a beam
undergoes focusing. A classical electromagnetic plane wave that
passes through a converging lens is no longer plane or
transversal~\cite{dorn:03,enk}. A substantial longitudinal field
component is present as well as the effect described above.

\section{ The effects of focusing}
In the analysis of the influence of a lens on quantum states  it
is important to bear in mind that it is incapable of turning a
pure state into a mixture unless information is lost. Indeed, pure
incoming state $|\Psi_{\rm in}\9$ transforms into a pure outcoming
state $|\Psi_{\rm out}\9$. The same information is encoded
differently in these two states, and the effective density matrix
captures our inability to access all of it. The lens action is
analyzed by mode matching. To achieve this, we find a mode
decomposition of solutions of the corresponding classical equation
(in the case of photons these are the electromagnetic wave
equations for the vector potential. Throughout this paper we use
the Coulomb gauge). With  each mode
$\bA_{\bk}=(2\pi)^{\frac{3}{2}}\bep^{\pm}_\bk e^{-i(\omega
t-\bk\cdot\bx)}$ a creation operator $\hat{a}^\dag_{\bk\pm}$ is
associated. The resulting quantum state
$|\bk,\pm\9=\hat{a}^\dag_{\bk\pm}|\vac\9 $ is normalized as
$\6\bk'\sigma'|\bk,\sigma\9=\delta^{(3)}(\bk-\bk')\delta_{\sigma
\sigma'}$.
All classical solutions are of the form
\beq
\bA(t,\bx)=A(t,\bx)\bal(t,\bx)=\int\!d\mu(\bk) \bal(\bk)A(\bk)e^{-i(\omega
t-\bk\cdot\bx)},
\label{classical}
\eeq
with $\bE=-\dot{\bA}$. The frequency $\omega=|\bk|c$, each field
component is split into the field strength $A(\bk)$ and the
transversal polarization part $\bal(\bk)$, $\bk\cdot\bal(\bk)=0$,
$|\bal(\bk)|=1$. The corresponding norm is
\beq
\|\bA\|=\|A\|=\left(\int\! d\mu(\bk)|A(\bk)|^2\right)^{\half}.
\label{normfix}
\eeq
 Therefore, a normalized positive energy solution of the form (\ref{classical})
 corresponds to a one-particle state
 \begin{eqnarray}
|\Psi\9=\sum_{\sigma=\pm}\int\! d\mu(\bk)\alpha_\sigma(\bk) f(\bk)
\hat{a}^\dag_{\bk\sigma}|\vac\9 \nonumber \\
=\int d\mu(\bk) f(\bk)|\bk,\bal(\bk)\9,\label{state}
\end{eqnarray}
where the coefficients $\alpha_\sigma$ are defined by
Eq.~(\ref{elliptic}), and $f(\bk)=A(\bk)/\|A\|$. The
transformation $|\Psi_{\rm in}\9\rightarrow |\Psi_{\rm out}\9$ is
obtained 
from the transformation of the incoming modes $\bA_{\bk}^{\rm in}$
into the outcoming modes $\bA_{\bk}^{\rm out}$ \cite{mandel,
qopt}. When the classical solution corresponding to the quantum
state is known, the effective polarization density matrix
Eq.~(\ref{reduce}) can be calculated as follows. We note that
\beq
\int A_n(\bx,t)A_m^*(\bx,t)d^3\bx=\int A_n(\bk)A_m^*(\bk)d\mu(\bk),
\eeq
for all $t$ and $n,m=x,y,z$. The density matrix (\ref{reduce}),
thus, can be written as
\beq
\rho=\frac{\int d\mu(\bk)\bA(\bk)\bA^\dag(\bk)}{\int d\mu(\bk)
|A(\bk)|^2}.\label{clden}
\eeq
We stress that this is a general expression  independent of
approximations that are generally used to calculate classical
solutions.

 The
analysis of the lens action should be performed in the vector
diffraction theory
\cite{born}, and  reliable estimates of a field in the focal
regions follow the techniques of Richards and Wolf
\cite{focal}. In some cases it is even possible to get exact solutions
of Maxwell equations \cite{enk}. Then, using Eq.~(\ref{clden}) we
find the outcoming quantum state whose effective reduced density
matrix is calculated according to Eq.~(\ref{reduce}).

To illustrate the importance of the longitudinal term, it is
sufficient to consider an incoming state with a definite momentum
and polarization $|\bk,\bep_\bk^\pm\9$, i.e. to approximate the
corresponding classical field by a plane monochromatic wave.  As
we showed above, this is a good approximation for  incoming
wavepackets that are the actually used in experiments. Moreover,
it is sufficient to use  ray tracing, while more refined
calculations should be used in conjunction with concrete
experimental schemes. Hence we take the outcoming classical field
as a spherical wave that converges to the geometric focus of a
thin lens with the focal length $f$, as illustrated on Fig.~1. The
polarization direction and the field strength at each point are
calculated using the eikonal equation and  ray tracing
\cite{born}.

The set-up is schematically presented on Fig.~1, while
calculations are given in  Appendix A. For the incoming states
$|\bk,\bep^\pm_\bk\9$, in the leading order the outgoing states
are
\beq
\rho_\pm\approx(1-\theta_\rmm^2/4)\left(\bay{ccc}
\half & \mbox{$\mp\frac{i}{4}$} & 0\\
\mbox{$\pm {i\over4}$} \ & \half \! & 0 \\
0 & 0 & 0\eay\right)+\theta_\rmm^2/4
\left(\bay{ccc} 0 & 0 & 0\\
0 & 0 & 0\\
0 & 0 & 1\eay\right), \label{rhoout}
\eeq
\begin{figure}[htbp]
\epsfxsize=0.48\textwidth
\centerline{\epsffile{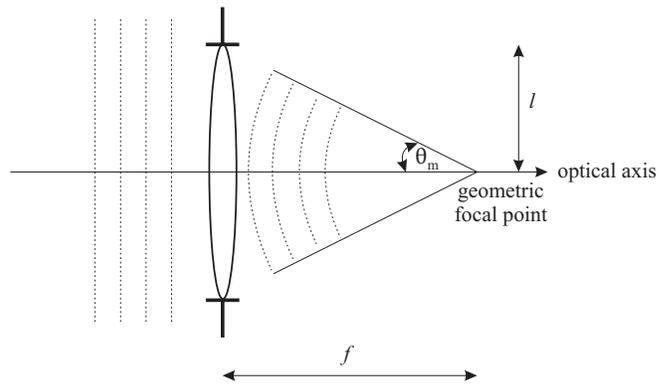}} \vspace*{-0.1cm} \caption{\small{Transformation
 of an incoming plane wave into a spherical
wave by the focusing system. }}
\end{figure}
This is the only  approximation which is consistent with use of
 Gaussian optics for  mode calculation.  All three eigenvectors are non-zero (even if,
of course, the one that is associated with an ideal  circular
polarization dominates).  The probability of error in
distinguishing between $\rho_+$ and $\rho_-$ is proportional to
the square of the numerical aperture,
\beq
P_E(\rho_+,\rho_-)=
\frac{\theta_{\rmm}^2}{8}.\label{probe}
\eeq

Since $\theta_{\rmm}\approx l/f$ where $l$ is the aperture radius,
for the moderate numerical aperture of  the order of $10^{-1}$,
 the probability of error gets to the level of percents,
compared with a case with no lens  present. The effects of
``longitudinal" polarization will be even more pronounced when
quantum states that correspond to classical doughnut-shaped
\cite{dorn:03} and other exotic modes are produced. This will be
enhanced by going to  higher numerical aperture, where the
longitudinal field can contain nearly 50\% of the total beam power
 in the limit ${\rm NA}\rightarrow 1$.

\section{Meaning of $\rho$}
 While the POVMs that were described in \cite{pt:03} are
legitimate theoretical constructions, the resulting $3\times 3$
table becomes experimentally relevant if the POVM corresponds to
some detection model. We establish this correspondence. Let us
compare predictions of the POVM $E_{x}$, $E_{y}$, $E_{z}$, $\sum
E_{j}=\1$ that gives the diagonal elements of the effective
density matrix $\rho_{jj}=\tr (E_{j}|\Phi\9\6\Phi|$)
\cite{pt:03,rmp} with the detection probabilities that can be obtained
from the first-order perturbational  calculations of the following
model detector. The operators $E_j$ are given explicitly in
Appendix B.

To facilitate the comparison we use an alternative form of the
diagonal elements of the density matrix as in Eq.~(\ref{clden}),
\beq
\rho_{jj}=W_j/W,\qquad j=x,y,z,
\eeq
where
\beq
 W_j=\int|\bA\cdot\hbj|^2dxdydz, \qquad
W=\sum W_j.
\eeq

 We discuss here  semiclassical detection
theory, since in the leading order the results of the full theory
agree with the semiclassical one \cite{mandel}. In the latter the
electromagnetic field is considered  classically, while the
photoelectrons are treated quantum-mechanically. The interaction
term is given by $\bp\cdot\bA$, where $\bA$ is a classical vector
potential (whose direction in the  polarization gauge is given by
the polarization vector $\bal$) and $\bp$ is the electron's
momentum operator. Accordingly, in the following we use a
classical language to describe the electromagnetic field. The area
$S$ of a planar detector is assumed to be much larger than the
cross section of the beam. We model the detector's sensitivity to
the wave's polarization by restricting the electron momentum to
lie only along a chosen direction.

Let us first consider a circularly polarized monochromatic beam in
the paraxial approximation
\cite{jac}. Assuming that it propagates along the $z$-axis,
we have
\beq
\bE(\bx,t)\approx\left(E(x,y)\bep^\pm+
\frac{i}{k}\left(\frac{\pad E}{\pad x}\pm i\frac{\pad E}{\pad x}\right)\hbz
\right)
e^{-i(\omega_0 t-k_0z)}, \label{epa}
\eeq
\beq
\bB\approx\mp i\bE, \label{bpa}
\eeq
where the basis polarization vectors are
$\bep^\pm=\sqrt{\half}(1,\pm i,0)$ and the beam radius $\tau$ is
much larger than a typical wavelength, $\tau k_0\gg 1$.

Let the  planar detector absorb the field along the $j$-axis
($j=x,y,z$) and  locate it at $z=z_0$. Hence,  electrons'
excitation rate is proportional to
 $\int |\bA(x,y,z_0)\cdot\hbj|^2dS$. Assuming a finite detection time $\Delta$ (or
a beam of finite duration $\Delta$, with the fields that are given
by the corresponding superpositions of
 $\bE$ and $\bB$ having different frequencies with some weight
 function $f=f(\omega-\omega_0)$), the excitation probability is proportional to
\beq
I_j=\int|\bA(x,y,z_0)\cdot\hbj|^2dSdt.
\eeq
Since the
detector is planar, we can change the integration variable from
$t$ to $z$ we get
\begin{eqnarray}
I_j=\Delta\int|\bA(x,y,z_0)\cdot\hbj|^2dxdy \noindent \\
=\int_{x,y}\int_{z=z_0-\Delta
c}^{z=z_0}\!\!|\bA(x,y,z)\cdot\hbj|^2dxdydz.
\end{eqnarray}
Hence $I_j=W_j$, and the diagonal elements of $\rho$ are indeed
related to the photocurrent as
\beq
\rho_{ss}=I_j/I, \qquad I=\sum I_j.
\eeq

Now consider a spherical wave which represents the EM field after
the lens, as in Eqs.~(\ref{knew})-(\ref{ax2}),(\ref{efild}). We
again consider a planar wave detector and detection time $\Delta$.
This time, however,  the normal to the detector plane is not
parallel to the Pointing vector, and thus $I_j/I$ are different
from $W_j/W$. Nevertheless, as will be seen shortly, these ratios
agree up to the order $\theta_\rmm^4$, while for the realistic
values of $\theta_\rmm$ the interesting effects (such as the
predicted error probability, Eq.~(\ref{probe})) are of the order
$\theta_\rmm^2$.

Let us assume that the detector is located at $z=z_0$ behind the
focus. We use  spherical coordinates with the origin in the focus
and $\theta=0$ at $z=z_0$, and  polar coordinates in the detector
plane. In this plane $\phi$ is the same as in the spherical
coordinates and $r=z_0
\tan\theta$. From the result of Appendix A, the field strength $E$ can be written as
\beq
E=E(R,\theta)=\frac{1}{(\cos\theta)^{3/2}}\frac{1}{r}=
\frac{e(\theta)}{r},
\eeq
and since we are working with a well defined frequency, a similar
decomposition is possible for $A$, i.e., $A=a(\theta)/r$. The
point $(\theta,\phi)$ on the detector plane is at a distance
$z_0/\cos\theta$ from the focus. The area element is
\beq
dS=\frac{\sin\theta}{\cos^3\theta}z_0^2d\theta d\phi,
\eeq
so the detection probability is proportional to
\begin{eqnarray}
I_j=\Delta
\int|a(\theta)|^2|\bal(\theta,\phi)\cdot\hbj)|^2\tan\theta
d\theta d\phi.
\end{eqnarray}
On the other hand, the integration over the shell $\Delta c$ gives
\beq
W_j=\Delta
\int|a(\theta)|^2|\bal(\theta,\phi)\cdot\hbj)|^2\sin\theta
d\theta d\phi.
\eeq
The normalized detection probabilities
\beq
p_j=I_j/I,
\eeq
differ therefore from the matrix elements of $\rho_{jj}=W_j/W$ by
the presence of additional factors $1/\cos\theta$ in each of the
integrals. However, when the results are expanded in terms of
$\theta_\rmm$, the difference between these two expressions is
only of the order of $\theta_\rmm^4$ or higher. Accordingly, the
leading order expansion for the probability of error,
Eq.~(\ref{probe}), remains the same  for the model detection
scheme that was described above. This discrepancy highlights the
fact that {\em different detection procedures lead to different
polarization density matrices}   \cite{aie}. Moreover, with a
slight abuse of the language, we note that the resulting effective
mixing is caused by the shape mismatch between the wave front and
the detector.

\section{Summary and outlook}
We have shown that an effective $3\times 3$ polarization density
matrix,  previously introduced on  formal grounds, indeed has a
direct experimental significance. It should be noted, however,
since there is no general polarization density matrix, different
detection procedures may lead to different effective
constructions.

 Presence of a significant longitudinal
part in the effective density matrix  imposes limits on detection
efficiency. The interaction between single atoms and
electromagnetic field, either classical or in a quantum coherent
state, is affected by focusing \cite{enk,enk:03}. The structure of
the effective density matrix shows that this will be the case also
for single photon states. Polarization-sensitive transitions in
the atoms will be suppressed by  strong focusing. For low
numerical apertures we expect these effects be proportional to its
square, reaching saturation on higher levels when $NA\rightarrow
1$.
 From the
point of view of quantum information theory, the fact that these
density matrices are inevitably mixed,  actually implies that
polarization qubits are always noisy. This intrinsic noise should
be taken into account in the analysis of the physical realizations
of quantum computing and in the security analysis of  quantum
cryptographic protocols. Similarly, sensitivity of  photon-atom
interaction to the focusing will affect the efficiency of
trapped-atom based quantum memory \cite{exam}.
\bigskip

\acknowledgments
\medskip
The work of NHL is supported by a grant from the Technion Graduate
School. Parts of this research were done during the visit of NHL
to the Perimeter Institute and of DRT to the University of
Queensland. The importance of focusing effects was pointed out to
us  by Eli Yablonovich. We thank Gerard Milburn, Petra Scudo,
Christine Silberhorn, Vlatko Vedral and Andrew White for useful
discussions and helpful comments.

\section*{APPENDIX A}

 Assume that the
optical axis is the $z$-axis, and approximate the incoming
classical wave as a plane wave with $\hbk=\hbz$  and polarization
$\bal$. Two systems of coordinates will be used in the following.
The plane $z=0$ is the plane where the lens is situated. In it we
define polar coordinates $(r,\phi)$, where $r$ is measured from
the optical axis in the $z=0$ plane. The unit vectors $\hbr,
\hbphi$ will always lie in the $z=0$ plane. The spherical
coordinates $(R,\theta,
\phi)$ are calculated from the focus (the angles $\phi$ in   both
coordinate systems are the same). The intersection between the
optical axis and $z=0$ plane has $r=f$ and $\theta=0$ coordinates,
while the lens aperture is bounded by $\theta_\rmm$.  The relation
\beq
 r=f \tan\theta,
\label{thetaofr}
\eeq
holds on the plane $z=0$, while  $\theta_\rmm$,
\beq
\tan \theta_\rmm=l/f,
\eeq
  is
determined by the nominal focal ratio \cite{born} $F=f/2l$, where
$l$ is the radius of the entrance pupil. In the Gaussian
approximation $\theta_\rmm$ equals
 to the numerical aperture.
 \begin{figure}[htbp]
\epsfxsize=0.46\textwidth
\centerline{\epsffile{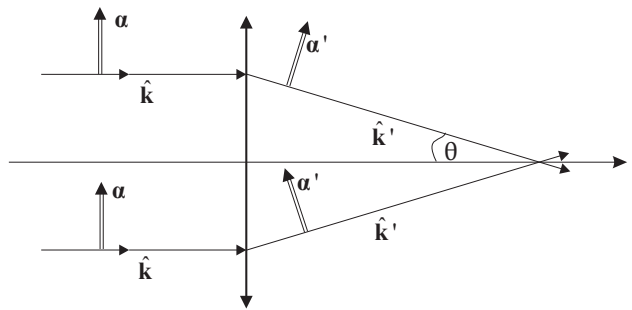}} \vspace*{-0.1cm} \caption{\small{Changes
 in a linear polarization according to the ray-tracing
model. The direction and polarization labels $\hbk$ and $\bal$ of
the ray are transformed into $\hbk'(\theta,\phi)$ and
$\bal'(\theta,\phi)$, respectively. }}
\end{figure}
Assuming that after passing through the lens the field is that of
a perfect spherical wave, the ray that hits the $z=0$ plane at
$(r,\phi)$ is deflected to the direction
\beq
\hbk'(\theta)=-\sin\theta\hbr+\cos\theta\hbz=
-\sin\theta(\cos\phi\hbx+\sin\phi\hby)+\cos\theta\hbz,
\label{knew}
\eeq
with $\theta$ given by Eq.~(\ref{thetaofr}). In passing through a
homogenious dielectric medium polarization directions are parallel
transported along each ray. At the boundaries the direction that
is parallel to the incidence plane is refracted to remain
transversal to the wave vector, while the perpendicular component
is unchanged
\cite{born,jac}. The incidence plane is spanned
 by the vectors $\hbz$ and $\hbr$, so that
the parallel and perpendicular components of polarization are
$\hbr$ and $\hbphi$, respectively. We  decompose the polarization
of the ray that passes through the $z=0$ plane at $(r,\phi)$, in
terms of $\hbr$ and $\hbphi$. For  linear $x$-polarization it is
given by
\beq
\bal_x\equiv \hbx=\cos\phi\hbr-\sin\phi\hbphi. \label{ax1}
\eeq
Having passed through the $z=0$ plane, the new direction of
polarization for the above ray becomes
\beq
\bal_x'(\theta,\phi)=\cos\phi(\cos\theta\hbr+\sin\theta\hbz)-\sin\phi\hbphi,
\label{ax2}
\eeq
were again $\theta$ is related to $r$ by Eq.~(\ref{thetaofr}). The
same calculation can be carried out for the linear $y$
polarization, giving
\beq
\bal_y'(\theta,\phi)=\sin\phi(\cos\theta\hbr+\sin\theta\hbz)+\cos\phi\hbphi.
\eeq
It is easy to see that  right and  left circular polarizations are
preserved up to a phase, $\bep^\pm_{\hbk}\rightarrow e^{\pm g
(\theta,\phi)}\bep^\pm_{\hbk'}$, where the precise form of $g$ is
irrelevant. Fig.~2 illustrates these changes.

 To complete the classical description of the
field as in Eq.~(\ref{classical}), we need the field strength
$E(\bk)$. Calculations of the intensity are based on the intensity
law of geometrical optics
\cite{born},
\beq
E^2dS=E'^2dS',
\eeq
where $E$, $dS$ and $E'$, $dS'$ are the field strength and the
area element at the respective wavefronts. Taking the initial
field strength to be unity, and considering that the wavefronts
before the lens are planar,
\beq
dS=2\pi r(\theta) dr(\theta)=2\pi
\frac{\sin\theta}{\cos^3\theta}f^2d\theta,
\eeq
and that after the lens they are spherical,
\beq
dS'=2\pi \sin\theta R^2 d\theta,
\eeq
 we get
\beq
E'(R,\theta)=\frac{1}{(\cos\theta)^{3/2}}\frac{f}{r}.
\label{efild}
\eeq
 Since both the plane
wave and the spherical wave are non-normalizable, we obtain the
density matrices from the two-dimensional integration over the
angular parts of the volume integral. For the incoming states
$|\bk,\bep^\pm_\bk\9$, taking into account that $\omega =|\bk
'|c={\rm const}$, the outgoing states are
\beq
\rho=\frac{\int_{\psi=0}^{2\pi}\!\int_{\theta=0}^{\theta_\rmm}
 \sin\theta d\theta d\psi
\bep_\pm(\theta,\psi)\bep_\pm^\dag(\theta,\psi)/\cos^3\theta}
{\int d\theta d\psi \sin\theta/\cos^3\theta},
\eeq
whose explicit form is given by Eq.~(\ref{rhoout}).

\section*{APPENDIX B}

The required POVM is introduced as follows.  The longitudinal
photons are used to define the necessary steps of our
construction. The POVM itself is build only with the physical
polarization states. Allowing for  longitudinal polarization makes
it possible to define a polarization state along an arbitrary
direction, say the $x$-axis, as
\beq
|\hbx\9=x_+(\bk)|\bep^+_\bk\9+x_-(\bk)|\bep^-_\bk\9+
x_\ell(\bk)|\bep^\ell_\bk\9,\label{decomp}
\eeq
where $x_\pm(\bk)=\bep^\pm_\bk\cdot\hbx$, and $x_\ell(\bk)=
\hbx\cdot\hbk$.
Note that $\6\hbx|\hby\9=\hbx\cdot\hby=0$, whence
\beq |\hbx\9\6\hbx|+|\hby\9\6\hby|+|\hbz\9\6\hbz|=\1\label{xyz}. \eeq

A projection operator that corresponds to the direction $\hbx$ is
\beq
P_{x}=|\hbx\9\6\hbx|\otimes \1_p=|\hbx\9\6\hbx|\otimes \int
d\mu(\bk)|\bk\9\6\bk|,
\eeq
where $\1_p$ is the unit operator in momentum space. The action of
$P_{x}$ on a physical state $|\Psi\9$ follows from
Eq.~(\ref{decomp}) and $\6\bep^\pm_\bk|\bep^\ell_\bk\9=0$. Only
the transversal part of $|\hbx\9$ appears in the expectation
value:
\beq
\6\Psi|P_{x}|\Psi\9=\int d\mu(\bk)|f(\bk)|^2|x_
 +(\bk)\alpha_+^*(\bk)+x_-(\bk)\alpha_-^*(\bk)|^2.
\eeq
Define the transversal part of $|\hbx\9$:
\begin{eqnarray}
|\bk,\bb_x(\bk)\9\equiv
 (|\bep^+_\bk\9\6\bep^+_\bk|+|\bep^-_\bk\9\6\bep^-_\bk|)|\hbx\9 \nonumber \\ =
 x_+(\bk)|\bep^+_\bk\9+x_-(\bk)|\bep^-_\bk\9,
\label{vector}
\end{eqnarray}
and likewise   $|\bb_y(\bk)\9$ and $|\bb_z(\bk)\9$. These three
vectors are neither of unit length nor mutually orthogonal.

Finally, a POVM element $E_{x}$ which is the physical part of
$P_{x}$, namely is equivalent to $P_{x}$ for physical states
(without longitudinal photons) is
\beq
E_{x}=\int d\mu(\bk)|\bk,\bb_x(\bk)\9\6\bk,\bb_x(\bk)|,
\eeq
and likewise for other directions. The operators $E_{x}$, $E_{y}$
and $E_{z}$ indeed form a POVM in the space of physical states,
owing to Eq.~(\ref{xyz}). It then follows from Eq.~(\ref{vector})
and similar definitions for the other directions that, for any
$\bk$,
\beq
|\bb_x(\bk)\9\6\bb_x(\bk)|+|\bb_y(\bk)\9\6\bb_y(\bk)|+
  |\bb_z(\bk)\9\6\bb_z(\bk)|={\1}_{\perp\bk},
\eeq
where ${\1}_{\perp\bk}$ is the identity operator in the subspace
of polarizations orthogonal to $\bk$.

\bigskip

\end{document}